\documentclass{aastex}
\usepackage{emulateapj5}
\lefthead{SHIGEYAMA \etal}
\righthead{Excavation of the first stars}

\def\etal{{et al. }}
\def\Msun{~M_{\odot} }
\def\cm3{{\rm ~cm}^{-3}}
\def\kms{{\rm ~km~s}^{-1}}
\def\ltsima{$\; \buildrel < \over \sim\;$}
\def\ltsim{\lower.5ex\hbox{\ltsima}}
\def\gtsima{$\; \buildrel > \over\sim \;$}
\def\gtsim{\lower.5ex\hbox{\gtsima}}

\received{2003 January 14}
\begin{document}
\title{Excavation of the first stars}

\author{Toshikazu Shigeyama$^{1}$, Takuji Tsujimoto $^{2}$, and Yuzuru Yoshii$^{1, 3}$}

\altaffiltext{1}{Research Center for the Early Universe, Graduate
School of Science, University of Tokyo, Bunkyo-ku, Tokyo 113-0033,
Japan; shigeyama@resceu.s.u-tokyo.ac.jp}

\altaffiltext{2}{National Astronomical Observatory, Mitaka-shi,
Tokyo 181-8588, Japan; taku.tsujimoto@nao.ac.jp}

\altaffiltext{3}{Institute of Astronomy, Graduate School of Science, University of Tokyo, Mitaka-shi, Tokyo, 181-0015, Japan; yoshii@ioa.s.u-tokyo.ac.jp}

\begin{abstract}
The external pollution of the first stars in the Galaxy is investigated. The first stars were born in  clouds composed of the pristine gas without heavy elements. These stars accreted gas polluted with heavy elements while they still remained in the  cloud.  As a result, it is found that they exhibit  a distribution with respect to the surface metallicity. We have derived the actual form of this distribution function. This metallicity distribution function strongly suggests that the recently discovered most metal-deficient star HE0107-5240 with [Fe/H]$=-5.3$ was born as a metal-free star and accreted gas polluted with heavy elements. Thus the heavy elements such as Fe in HE0107-5240 must have been supplied from supernovae of later generations exploding inside the  cloud in which the star had been formed. The elemental abundance pattern on  the surface of stars suffering from such an external pollution should not be diverse but exhibit the average pattern of numerous supernovae. Future observations for a number of metal-deficient stars with [Fe/H]$<-5$ will be able to prove or disprove this external pollution scenario. Other possibilities to produce a star with this metallicity are also discussed. 
\end{abstract}

\section{INTRODUCTION}
Though the first generation stars (so called Population III stars) in the Galaxy must have been formed from the pristine gas with no heavy elements, all the stars observed so far contain some amount of heavy elements such as iron (Fe). According to our scenario, the recent discovery of the most metal-deficient star HE0107-5240 with an Fe abundance as low as $1/200,000$ of the solar value \citep{Christlieb_02} suggests that Population III (Pop III) stars contained long-lived low-mass objects \citep{Yoshii_79}. Thus there have survived Pop III stars even in the present Milky Way. To identify such stars, we have investigated how the surface metallicity of a Pop III star increases by accreting gas polluted with heavy elements and found that HE0107-5240 is much more likely to have been formed from the pristine gas cloud than from a gas cloud that had already contained Fe \citep[see also][in the case of globular clusters and field stars]{Thoul_02}.

No stars without heavy elements have been discovered to date probably because stars experience the external pollution. Due to their gravity, stars can accrete the interstellar matter (ISM) polluted with heavy elements. A star with a mass $M$ moving at a velocity $v$ relative to gas accretes gas at a rate $\dot{M}(v)=1.120\pi\left(2GM/v^2\right)^2\rho v$ \citep{Bondi_52}, where  $\rho$ denotes the density of the gas and $G$ the gravitational constant. Thus a long-lived star moving at a velocity of  $\sim 200\kms$ due to the gravitational potential of the Milky Way accretes gas at a rate of the order of  $10^{-17} \Msun\,{\rm yr}^{-1}$. The ISM has been enriched with heavy elements since the first stars were formed. Hence this accretion may increase the surface metallicity to an observable level if the surface convective layer is as shallow as in a metal-deficient dwarf star on the main sequence of the stellar evolution \citep{Yoshii_81}. On the other hand, a metal-deficient red-giant star would reduce the external pollution to a negligible level by diluting the accreted heavy elements inside its more extended surface convective region \citep{Fujimoto_95, Fujimoto_00}. The mass of the surface convective layer in a red-giant star becomes about two orders of magnitude larger than that on the main sequence. Therefore the recently discovered most metal-deficient red-giant star HE0107-5240 has not increased its surface metallicity only by accretion but has reached such a low value through an accretion followed by the dilution due to the growing of the convective envelope.

Since the accretion rate is inversely proportional to $v^3$, a much more rapid accretion can be expected when a star remains in its birth place, that is, in a  cloud where a star moves with a smaller velocity relative to gas. The minimum mass that can gravitationally collapse immediately after the recombination of the universe indicates a typical mass of the first clouds to be about $10^6\Msun$ \citep{Peebles_68}. After contraction, a cloud formed a self-gravitating system in an equilibrium configuration with a velocity dispersion $\sigma\sim 10\kms$  corresponding to the sound speed at the recombination. Thus stars in this cloud move at  $\sqrt{2}\sigma\sim14\kms$ relative to the turbulent gas with the same velocity dispersion and accrete gas at a rate $\sim$8,000 times larger than those moving at a velocity of the order of 200 $\kms$ with respect to the surrounding ISM. The density of the cloud being greater than the mean density of the Galaxy, the accretion rate is also enhanced, whereas the lifetime $t_{\rm life}$ ($<$ 1 Gyr) of a cloud being shorter than the age of the Galaxy, the accreted mass might be reduced by a factor of $\sim$20.
A quantitative description of the external pollution by accretion will be presented in the next section. 

\begin{figure*}[ht]
\begin{center}
\plotone{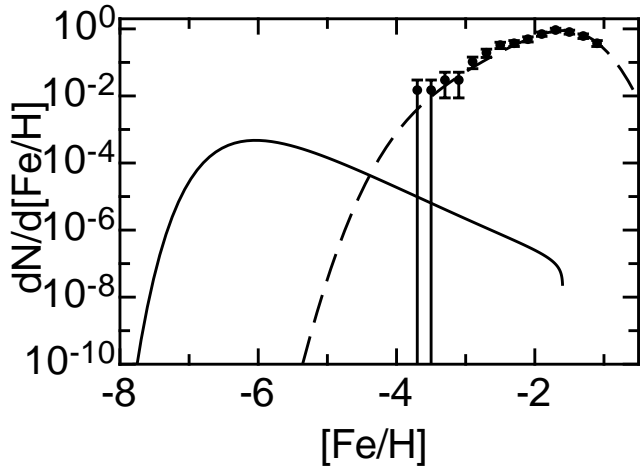}
\caption{The metallicity distribution functions (MDFs) for Pop III (solid line) and II (dashed line) stars with respect to [Fe/H]. The functions are normalized to unity when they are integrated over the whole range of the metallicity. The points are the MDF for Pop II dwarf stars \protect{\citep{Ryan_91}}.}
\end{center}
\end{figure*}
\section{Accretion onto metal-free stars}
From the virial theorem, the average density $\rho$ of the gas is given by the formula $\rho=3\sigma^6\big/\left(4\pi G^3M^2\right)$. The accretion rate is assumed constant during the lifetime of the cloud for simplicity. A star thus increases its surface abundance of Fe according to the formula
 \begin{eqnarray}
{\rm [Fe/H]}=&\log\left\{\dot{M}(v)\int_0^{t_{\rm life}}dtX_{\rm Fe}(t)/\left(X_{\rm H}M_{\rm mix}\right)\right\}& \nonumber\\
&-\log\left(X_{\rm Fe}\big/X_{\rm H}\right)_\odot.&
\end{eqnarray}
Here [Fe/H] denotes the logarithmic value of the ratio of Fe to hydrogen (H) relative to the solar value $\left(X_{\rm Fe}\big/X_{\rm H}\right)_\odot$,  $X_{\rm Fe}(t)$ denotes the Fe abundance of the gas at time $t$, $X_{\rm H}$  and $M_{\rm mix}$ denote the H abundance (=0.75) and mass of the surface convection zone in a star, respectively. Since the accretion rate is the only factor that depends on the velocity $v$ relative to gas, the number of Pop III stars per unit interval of [Fe/H] (metallicity distribution function, MDF) can be expressed using the distribution function $f(v)$ with respect to velocity as
 \begin{equation}
\frac{dN}{d{\rm [Fe/H]}}\left({\rm [Fe/H]}\right)=f(v)\left|\frac{dv}{d{\rm [Fe/H]}}\right|=\frac{\ln10}{3}vf(v).
\end{equation}
Here stars and gas are assumed to follow the Gaussian distribution function with the same velocity dispersion $\sigma=10\kms$. Eliminating $v$ from equations (1) and (2) yields the MDF.

To describe the evolution of Fe abundance $X_{\rm Fe}(t)$ of the gas, we use a chemical evolution model by \citet{Tsujimoto_99} that successfully reproduces the distribution of halo field stars with respect to their Fe abundances [Fe/H]  observed by \citet{Ryan_91} . This model assumes that stars are formed from the gas compressed by the blast wave generated by a single supernova (SN) and inherit the elemental abundance pattern of the compressed gas comprising the ejecta of the SN. It is also assumed that the star formation was initiated by supernova explosions of Pop III massive stars. The star formation ends when SN remnants overlap each other to heat up the cloud. Since a majority of SNe eject roughly 0.1$\Msun$ of Fe per event \citep[e.g.,][]{Hamuy_03} and the mass of the ISM compressed by a single SN is of the order of 10$^5\,\Msun$ \citep{Shigeyama_98}, it is predicted that most stars formed in the compressed ISM will have a metallicity of [Fe/H]$>-4$ \citep{Tsujimoto_99b}. To achieve such low values, the accretion process onto metal-free stars must be taken into account. This is consistent with the observed MDF of halo field stars belonging to Pop II as shown in Figure 1. A similar scenario has been proposed for the formation of globular clusters and Pop II field stars \citep{Parmentier_99}. As mentioned above, our model requires the existence of Pop III stars in a gravitationally bound cloud. The recently discovered most metal-deficient star suggests that these Pop III stars include some stars with lifetimes as long as the age of the universe. Thus the external pollution of these stars must influence the current MDF of extremely metal-deficient stars. In this model, the mass fraction $x$ of Pop III stars is a free parameter to determine the duration of the star formation or the lifetime of the cloud. For a given mass of the cloud, this fraction must be greater than a certain value to form at least one massive star leading to SN explosion. A cloud mass of $10^6\Msun$ requires $x\geq 10^{-4}$ that leads to $t_{\rm life}\leq 700$ Myr \citep{Tsujimoto_99b}. Here the initial mass function (IMF) is assumed to be the same as the one derived for the solar neighbourhood \citep{Salpeter_55, Tsujimoto_97}. 

The mass of HE0107-5240 is estimated to be $0.8\Msun$ \citep{Christlieb_02}. Theoretical stellar evolution models suggest that the mass range of stars that were formed first in the Galaxy and have now evolved to red-giant stars must be quite narrow around 0.8$\Msun$ because the red-giant phase is brief as compared to the whole lifetime. Thus the MDF only for 0.8$\Msun$ stars is considered in the following. Theoretical calculations \citep[e.g.,][]{Fujimoto_00b} show that a Pop III giant star with this mass has a surface convection zone with $M_{\rm mix}=0.35\Msun$ at present. To illustrate the maximum effect of the accretion onto 0.8$\Msun$ stars, we have calculated the MDF of Pop III stars with this minimum fraction $x\,(=10^{-4})$ corresponding to the longest lifetime of a cloud. The result is shown in Figure 1 together with the MDF for later generation stars \citep{Tsujimoto_99b}. These MDFs have been normalized to unity when they are integrated over the whole range of metallicity. Here the IMF is assumed not to vary with time during the halo star formation. Figure 2 shows the relation between $v$ and [Fe/H] given by Equation (1). A larger fraction $x$ of Pop III stars would lead to a shorter lifetime of the cloud. As a consequence, the accreted mass would be reduced. Thus the maximum of the MDF for Pop III stars would be shifted toward a smaller [Fe/H] without changing the shape of the distribution. Of course, the MDF for Pop III stars would increase relative to that for Pop II stars with increasing $x$. Therefore Figure 1 strongly suggests that most of stars with [Fe/H]$=-5.3$ such as HE0107-5240 belong to Pop III. Since the shape of the predicted distribution is invariant to the parameter $x$, future observations will be able to test the accretion model described above. This external pollution inevitably converges the elemental abundance pattern on the stellar surface to the uniform one represented by the average yield from SNe with all possible progenitor masses. The observed pattern in HE0107-5240 \citep{Christlieb_02} is compatible with the average pattern for halo field stars, although this uniformity in chemical composition is questionable \citep{Jehin_99}.  However, the observed enhancement of the carbon abundance ([C/H]$=-1.3$) is ascribed to a dredge-up of C and nitrogen to the surface taking place in an extremely metal-poor star such as HE0107-5240 \citep{Fujimoto_00b, Christlieb_02} and cannot be explained by supernovae before the birth of this star, because a supernova supplies not only C but also oxygen, which was not observed.

The short lifetime of a cloud indicates that stars were accreting gas while they were still unevolved dwarf stars. Hence a dwarf star born from the pristine gas should have exhibited a fairly high metallicity when the cloud was disrupted by supernova explosions at time $t_{\rm life}$. Then the star evolved off the main sequence to become a red-giant star if the stellar mass is as massive as 0.8 $\Msun$. In the mean time, the surface convection zone grew and diluted the accreted heavy elements. As a result, the surface metalliciy has decreased by about two orders of magnitudes to be $\sim 1/100,000$ of the solar value. On the other hand, less massive Pop III stars still remain on the main sequence with a metallicity of $\sim 1/1,000$ of the solar value. 

A strong stellar wind may prevent a star from accreting the ISM. However, even if a main sequence star blew a wind as strong as the sun, it would hardly stop the accretion because the typical accretion rate considered here is  $2\times 10^{-13}\Msun\,{\rm yr}^{-1}$ (Fig. 2) that is significantly larger than the solar mass loss rate of  $2.5\times 10^{-14}\Msun\,{\rm yr}^{-1}$ \citep{Cox_00}. Furthermore, a metal-free dwarf star with a mass of 0.8$\Msun$ has a surface convection zone much shallower than the sun leading to a less active corona in HE0107-5240 on the main sequence. As a result, it is likely that the wind from HE0107-5240 was much weaker than the sun while this star was accreting gas.

\begin{figure*}[ht]
\begin{center}
\plotone{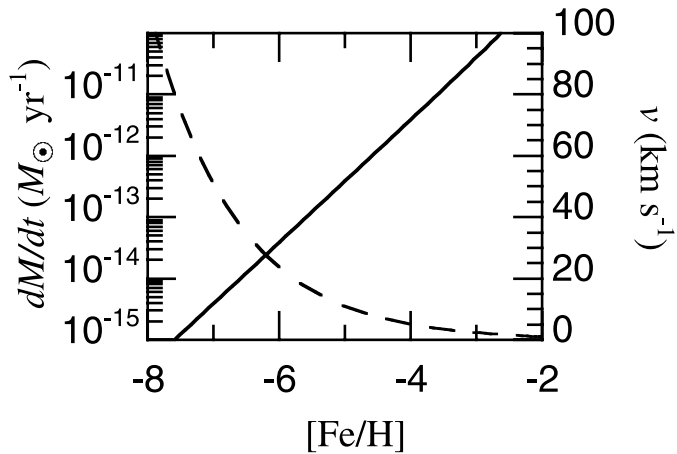}
\caption{The mass accretion rate (solid line; in the unit of $\Msun\,{\rm yr}^{-1}$) and the velocity of a star relative to gas (dashed line) as functions of the surface metallicity  [Fe/H] of Pop III star with a surface convection zone of mass $M_{\rm mix}=0.35\Msun$.}
\end{center}
\end{figure*}

\section{Alternative origins of heavy elements}
Other possibilities to produce a star with this metallicity will be discussed in this section.

\subsection{External pollution by a single supernova}
If a low mass Pop III star was formed as a member of a binary system with a massive companion, the companion star might have injected a part of the ejecta into the atmosphere of the dwarf star when it exploded as a SN. Since the average speed of ejecta is $\sim 3,000\kms$, the mass injection into the dwarf star is expected to be extremely inefficient. If the ejecta that expand into the geometrical cross section of the star with the radius $R$ are injected, the total mass $M_{\rm inj}$ of the injected ejecta will be given by the formula $M_{\rm inj}=M_{\rm ej}\left(R\big/2D\right)^2$, where $M_{\rm ej}$ denotes the total mass of the ejecta and $D$ the separation of the stars. This will give the maximum injected mass because the ejecta move at a velocity much greater than the escape velocity of the star. To explain the observed value of [Fe/H]$=-5.3$ on the surface of a star with the solar radius, the separation at the time of SN explosion must be closer than $\sim 10^{13}$ cm if the SN ejected $\sim 0.1\Msun$ of Fe. In reality, the ejecta with sufficiently high velocities cannot be injected into the star even if they initially proceed toward the star. Since the radius of a red super-giant star could exceed this separation, these two stars would merge into a single massive star before SN explosion.

If the companion star was so massive ($\gtsim 100\Msun$) that it underwent a so-called pair instability supernova that ejected more than 10$\Msun$ Fe \citep[e.g.,][]{Heger_02}, it was able to inject enough Fe into the dwarf star to explain the observed metallicity even if the separation was as large as $\sim 10^{14}$ cm. According to current theoretical models for this type of supernovae \citep{Heger_02}, however, it seems that no SN can inject heavy elements with the abundance pattern observed in HE0107-5240. Since different elements expand at different velocities in SN ejecta, detailed hydrodynamic calculations are needed to explore the abundance pattern of heavy elements injected by this process. A pair instability supernova is so energetic that it can disrupt a  cloud with a mass of $10^6\Msun$. As a result, other Pop III stars inside the same cloud would not accrete the gas. Pair instability supernovae might prevent most of metal-free stars from an external pollution.

\subsection{HE0107-5240 as a Pop II star}
There might still be a possibility that stars with [Fe/H]$=-5.3$ were formed from gas into which the precursor SNe had already injected heavy elements, that is, belong to Pop II. It has been suggested that the elemental abundance pattern of a metal-deficient star (with [Fe/H]$\ltsim -2$) was imprinted by a single SN and coincides with that in the gas eventually swept up by the blast wave generated from this SN \citep{Shigeyama_98}. Thus the amount of Fe corresponding to [Fe/H]$=-5.3$ suggests that the precursor SN ejected only $\sim 0.0005\Msun$ of Fe. In fact, the current SN observations have revealed that each of the least luminous SNe ever observed, SN 1997D and SN 1999br, ejected a small amount of Fe, while a majority of SNe eject an amount of Fe per event at least one order of magnitude larger \citep[see][]{Hamuy_03, Zampieri_03}. Theoretical SN models suggest that the deficiency of Fe must be a result of the fall back of some Fe in the ejecta onto the compact remnant at the center \citep{Turatto_98}. As a consequence, the ejecta exhibit an apparent enhancement of $\alpha$-elements such as magnesium (Mg) relative to Fe. By contrast, the observed ratio in HE0107-5240, [Mg/Fe]$=0.2$, is comparable with the average value for halo field stars. Therefore it is unlikely that HE0107-5240 inherited heavy elements ejected by a subluminous SN. On the contrary, two extremely metal-deficient stars CS22949-037 and CS29498-043 exhibit distinct enhancements of $\alpha$-elements with respect to Fe such as [Mg/Fe]$>1$ \citep{Depagne_02, Aoki_02}. These stars might have inherited heavy elements from subluminous SNe \citep{Tsujimoto_03}.

There is, of course, inhomogeneity in the matter swept up by a single SN. Thus it could be expected that stars with [Fe/H]=-5.3 were formed even from luminous SNe. A numerical study of metal enrichment by the first SN in an inhomogeneous ISM apparently denies this possibility. 3-D hydrodynamic calculations of a SN explosion at different sites has shown that there is little gas with [Fe/H]=-5.3 inside the matter swept up by the SN \citep{Nakasato_00}. In addition, the MDF must be easily distinguished from that presented in Figure 1 because the mechanism to acquire Fe are quite different from the accretion. The MDF for stars with [Fe/H]$<-5$ is indispensable for identifying the origin of heavy elements in these stars. Further differences must exist in the abundance pattern of each star that reflects the precursor SN and exhibit a feature different from each other. This is in contrast with the consequence of the accretion mentioned above.

\section{Conclusions}
Here we have discussed the gas accretion onto metal-free stars while they stayed in a  cloud with a mass of $\sim 10^6\Msun$ and predicted the MDF for Pop III stars as a consequence of the external pollution. After the disruption of the  cloud, stars have been moving at  $\sim 200\kms$  under the gravitational potential of the whole galaxy and no longer accreting a substantial amount of matter. This MDF strongly suggests that the recently discovered most metal-deficient star HE0107-5240 belongs to Pop III. The expected elemental abundance patterns on the surfaces of these Pop III stars with external pollution should not be diverse but exhibit the average pattern of numerous SN explosions. The observed pattern in HE0107-5240 satisfies this requirement. The argument presented here can be tested by future observations for a number of extremely metal-deficient stars with [Fe/H]$<-5$.

\acknowledgements
This work was supported in part by a Grant-in-Aid for Scientific
Research (11640229, 12640242) from the Ministry of Education, Culture,
Sports, Science, and Technology of Japan.


\begin{thebibliography}{}
\bibitem[Aoki et al.(2002)]{Aoki_02}
Aoki, W., Norris, J. E., Ryan, S. G., Beers, T. C. \& Ando, H. 2002, \apj, 576, L141
\bibitem[Bondi(1952)]{Bondi_52}
Bondi, H. 1952, \mnras, 112, 195
\bibitem[Christlieb et al.(2002)]{Christlieb_02}
Christlieb, N. et al. 2002, \nat, 419, 904
\bibitem[Cox(2000)]{Cox_00}
Cox, A. N. 2000, Allenfs Astrophysical Quantities Fourth Edition. Springer-Verlag New York
\bibitem[Depagne et al.(2002)]{Depagne_02}
Depagne, E. et al. 2002, \aap,  390, 187
\bibitem[Fujimoto et al.(1995)]{Fujimoto_95}
Fujimoto, M. Y., Sugiyama, K., Iben, I. Jr. \& Hollowell, D. 1995, \apj,  444, 175
\bibitem[Fujimoto, Ikeda, \& Iben(2000)]{Fujimoto_00}
Fujimoto, M. Y., Ikeda, Y. \& Iben, I. Jr. 2000, \apj, 529, L25
\bibitem[Hamuy(2003)]{Hamuy_03}
Hamuy, M. 2003, \apj,  582, 905
\bibitem[Heger \& Woosley(2002)]{Heger_02}
Heger, A. \& Woosley, S. E. 2002, \apj, 567, 532
\bibitem[Jehin et al.(1999)]{Jehin_99}
Jehin, E., Magain, P.,  Neuforge, C., Noels, A., Parmentier, G., \& Thoul, A.~A. 1999, \aap, 341, 
241 
\bibitem[Nakasato \& Shigeyama(2000)]{Nakasato_00}
Nakasato, N. \& Shigeyama, T. 2000, \apj, 541, L59
\bibitem[Parmentier et al.(1999)]{Parmentier_99}
Parmentier, G.,  Jehin, E., Magain, P., Neuforge, C., Noels, A., \& Thoul, A.~A. 1999, 
\aap, 352, 138 
\bibitem[Peebles \& Dicke(1968)]{Peebles_68}
Peebles, P. J. E., \& Dicke, R. H. 1968, \apj, 154, 891
\bibitem[Ryan \& Norris(1991)]{Ryan_91}
 Ryan, S. G. \& Norris, J. E. 1991, \aj, 101, 1835
\bibitem[Salpeter(1955)]{Salpeter_55}
Salpeter, E., E. 1955, \apj, 121, 161
\bibitem[Shigeyama \& Tsujimoto(1998)]{Shigeyama_98}
Shigeyama, T. \& Tsujimoto, T. 1998, \apj, 507, L135
\bibitem[Thoul et al.(2002)]{Thoul_02}
Thoul, A., Jorissen, A., Goriely, S., Jehin, E., Magain, P., Noels, A., \& Parmentier, G. 2002, \aap, 383, 491 
\bibitem[Tsujimoto et al.(1997)]{Tsujimoto_97}
Tsujimoto, T., Yoshii, Y., Nomoto, K., Matteucci, F., Thielemann,
F.-K., \& Hashimoto, M. 1997, \apj, 483, 228
\bibitem[Tsujimoto, Shigeyama, \& Yoshii(1999)]{Tsujimoto_99}
Tsujimoto, T., Shigeyama, T. \& Yoshii, Y. 1999, \apj, 519, L63
\bibitem[Tsujimoto \& Shigeyama(2003)]{Tsujimoto_03}
Tsujimoto, T., \& Shigeyama, T. 2003, \apj, 584, L87
\bibitem[Turatto et al.(1998)]{Turatto_98}
Turatto, M. et al. 1998, \apj, 498, L129
\bibitem[Yoshii \& Sabano(1979)]{Yoshii_79}
Yoshii, Y. \& Sabano, Y. 1979, \pasj, 31, 505
\bibitem[Yoshii(1981)]{Yoshii_81}
Yoshii, Y. 1981, \aap, 97, 280
\bibitem[Zampieri et al.(2003)]{Zampieri_03}
Zampieri, L., et al. 2003, \mnras, 338, 711
\bibitem[Tsujimoto et al.(1999)]{Tsujimoto_99b}
\bibitem[Fujimoto et al.(2000)]{Fujimoto_00b}
\end{thebibliography}
\end{document}